\documentclass[twocolumn,prb,superscriptaddress,longbibliography]{revtex4-2}

\usepackage{amsmath,amsfonts,amssymb}
\usepackage{graphicx,color}
\usepackage[breaklinks=true]{hyperref}
\usepackage{breakcites}
\usepackage{siunitx}
\usepackage{epstopdf}
\usepackage{float}
\usepackage{multirow}
\usepackage{hhline}
\usepackage{ulem}
\usepackage{mathtools}
\usepackage[caption=false]{subfig}
\usepackage{soul}
\hypersetup{hidelinks,colorlinks=true,allcolors=BLUE}
\usepackage{physics}
\usepackage{nicefrac}
\usepackage{bbold}
\makeatletter

\@ifundefined{textcolor}{}
{%
 \definecolor{BLACK}{gray}{0}
 \definecolor{WHITE}{gray}{1}
 \definecolor{RED}{rgb}{1,0,0}
 \definecolor{GREEN}{rgb}{0,1,0}
 \definecolor{BLUE}{rgb}{0,0,1}
 \definecolor{CYAN}{cmyk}{1,0,0,0}
 \definecolor{MAGENTA}{cmyk}{0,1,0,0}
 \definecolor{YELLOW}{cmyk}{0,0,1,0}
}

\makeatother

\begin{document}

\title{Exploring Ququart Computation on a Transmon using Optimal Control}

\author{Lennart Maximilian Seifert}
\affiliation{Department of Computer Science, University of Chicago, Chicago, Illinois 60637, USA}

\author{Ziqian Li}
\affiliation{James Franck Institute, University of Chicago, Chicago, Illinois 60637, USA}
\affiliation{Department of Physics, University of Chicago, Chicago, Illinois 60637, USA}
\affiliation{Department of Applied Physics, Stanford University, Stanford, California 94305, USA}

% \author{Jason Chadwick}
% \affiliation{UChicago}

\author{Tanay Roy}
\affiliation{James Franck Institute, University of Chicago, Chicago, Illinois 60637, USA}
\affiliation{Department of Physics, University of Chicago, Chicago, Illinois 60637, USA}

\author{David I. Schuster}
\affiliation{James Franck Institute, University of Chicago, Chicago, Illinois 60637, USA}
\affiliation{Department of Physics, University of Chicago, Chicago, Illinois 60637, USA}
\affiliation{Department of Applied Physics, Stanford University, Stanford, California 94305, USA}
\affiliation{Pritzker School of Molecular Engineering, University of Chicago, Chicago, Illinois 60637, USA}

\author{Frederic T. Chong}
\affiliation{Department of Computer Science, University of Chicago, Chicago, Illinois 60637, USA}

\author{Jonathan M. Baker}
\affiliation{Department of Computer Science, University of Chicago, Chicago, Illinois 60637, USA}
\affiliation{Duke Quantum Center, Duke University, Durham, North Carolina 27701, USA}

\date{\today}

\begin{abstract}
% Contemporary quantum computers encode and process quantum information in binary qubits ($d=2$).
% 
% However, in many architectures, higher energy levels are unused computational resources.
Contemporary quantum computers encode and process quantum information in binary qubits ($d=2$). However, many architectures include higher energy levels that are left as unused computational resources.
% 
% We demonstrate a superconducting ququart ($d=4$) processor and present the application of quantum optimal control together with efficient gate decompositions to implement high-fidelity ququart gates. 
We demonstrate a superconducting ququart ($d=4$) processor and combine quantum optimal control with efficient gate decompositions to implement high-fidelity ququart gates. 
%
% We distinguish between viewing the ququart as a generalized four-level qubit and as an encoded pair of qubits, and characterize the gates independently.
We distinguish between viewing the ququart as a generalized four-level qubit and an encoded pair of qubits, and characterize the resulting gates in each case.
% 
% In randomized benchmarking (RB) experiments, we observe gate fidelities $\geq 95\%$ using optimal control and identify coherence as the primary limiting factor. 
In randomized benchmarking experiments we observe gate fidelities $\geq 95\%$ and identify coherence as the primary limiting factor. 
Our results validate ququarts as a viable tool for quantum information processing.
\end{abstract}

\maketitle

\section{Introduction}
While current efforts to build quantum computers mostly focus on the fabrication and high-fidelity control of two-level qubits, many proposed implementations like superconducting transmons or trapped ions provide a much larger Hilbert space with more energy levels present. Their influence is typically aimed to be suppressed through careful device engineering, however, these states are readily available for computation.

Qudits, the $d$-level generalization of qubits, have gained a lot of interest as they provide an exponential increase in Hilbert space dimension ($d^N$ versus $2^N$), allowing for the adaption and simplification of a variety of algorithms~\cite{2020qudit, nikolaeva2022efficient, deller2022quantum, bocharov_factoring_2017, nguyen_quantum_2019}. Extensive research has been conducted for the qutrit case $d=3$: While theoretical studies have shown benefits for quantum compilation~\cite{baker_efficient_2020, baker_improved_2020} and improved schemes for quantum error correction~\cite{2016VSLQ, li2023autonomous, muralidharan_overcoming_2017, majumdar_quantum_2018}, experimental implementations of qutrits have been presented on several architectures~\cite{blok_quantum_2021, ringbauer_a_2022, hrmo_native_2022, luo_quantum_2019, morvan_qutrit_2021,  goss_high-fidelity_2022, 2021qutritalg, roy2022realization, 2023qutritcoupler, li2023hardware}.
%, demonstrating multi-qutrit gates and entire algorithms~\cite{morvan_qutrit_2021,  goss_high-fidelity_2022, 2021qutritalg, roy2022realization, 2023qutritcoupler, li2023hardware}.

Ququarts ($d=4$) can be explored from a similar perspective, and they furthermore offer the alternative interpretation of storing the information content of two qubits. Theoretical works have looked at the encoding of qubit pairs into ququarts to study advantages for quantum circuit compilation~\cite{litteken_qompress:_2023, litteken_dancing_2023}. Increased hardware utilization, efficient internal two-qubit gates and reduced routing costs render this design highly promising. Recent experimental results have shown the efficient realization of a variational quantum algorithm~\cite{cao_emulating_2023} and the verification of the entropic inequality~\cite{dong_simulation_2022} under this scheme. 

Quantum optimal control comprises a set of methods to find hardware-specific control pulses to make a quantum system undergo a desired transformation. A variety of frameworks have been developed~\cite{petersson_optimal_2021, gunther_quandary:_2021, goerz_krotov:_2019, khaneja_optimal_2005, propson_robust_2021, q-ctrl_boulder_2023, ball_software_2021, wu_data-driven_2018, egger_qiskit_2021} and applied in both theory and experiment to solve problems like state transfer~\cite{gunther_quantum_2021, goldschmidt_model_2022, meitei_gate-free_2021a} and the implementations of gates~\cite{wu_high-fidelity_2020, seifert_time-efficient_2022, chadwick_efficient_2023, carvalho_error-robust_2021}.

In this work, we study the realization of single-ququart operations on a superconducting transmon under both the qudit perspective as well as the encoded-qubit perspective. We compare two approaches for each gate: An optimized decomposition into natively supported gates on our system, and direct implementation through quantum optimal control. Using the frameworks Juqbox~\cite{petersson_optimal_2021} and Boulder Opal~\cite{q-ctrl_boulder_2023}, we find optimal control pulses that drive all three transitions in parallel while respecting hardware constraints. We perform quantum process tomography (QPT) as well as randomized benchmarking (RB) to characterize the fidelity of different gate implementations, where we distinguish between the ququart Clifford group $\mathcal{C}_4$ and the two-qubit Clifford group $\mathcal{C}_2^{\otimes 2}$. We focus on the ququart Hadamard $H_4$ and the two-qubit Hadamard tensor product $H \otimes H$ due to their similar structure, and include further results in the appendix. To the best of our knowledge, our work is the first to apply optimal control to manipulate a transmon ququart and study the differences in ququart RB schemes.

\begin{figure*}[htbp]
    \centering
    \includegraphics[width=\linewidth]{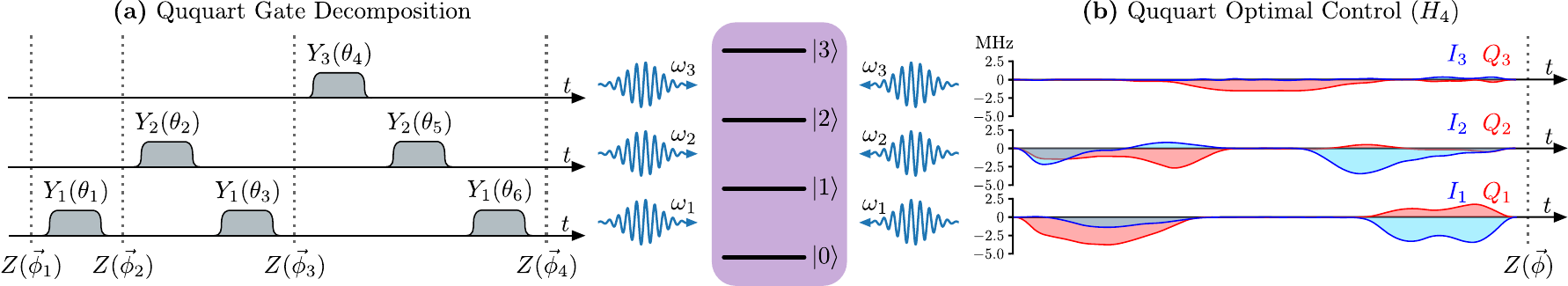}
    \caption{Two ways of realizing arbitrary quantum gates on a four-level ququart, which is controlled by three modulated carrier waves with resonant frequencies $\omega_j$. (a) Any ququart gate can be decomposed into a series of at most six qubit-like $Y$ rotations interleaved with ququart phase gates. Given a target unitary, the parameters $\theta_i$ and $\vec{\phi}_i$ can be calculated numerically. (b) Lifting the gate abstraction, quantum optimal control can be used to determine a control signal that operates on all subspaces in parallel and manipulate the entire ququart state. A final virtual $Z$ gate adjusts the relative phases. The shown pulses together with $\vec{\phi} [\mathrm{rad}] =(4.265, 4.590, 0.000)$ implement a ququart Hadamard $H_4$ in $T = 350 \, \mathrm{ns}$.} 
    \label{fig:decomp_pulse}
\end{figure*}

\section{One ququart or two qubits}
The concept of the ququart extends the computational unit of a qubit by two levels and describes the special qudit case $d=4$. Single-ququart gates are then given by $4 \times 4$ unitary matrices and some of them can thus be seen as generalizations of familiar single-qubit gates~\cite{gottesman_fault-tolerant_1999}:
\begin{equation}
\begin{gathered}
    X_4 = \mqty[
                0 & 0 & 0 & 1 \\
                1 & 0 & 0 & 0 \\
                0 & 1 & 0 & 0 \\
                0 & 0 & 1 & 0 
            ], \quad
    H_4 = \frac{1}{2}\mqty[
                1 & 1 & 1 & 1 \\
                1 & i & -1 & -i \\
                1 & -1 & 1 & -1 \\
                1 & -i & -1 & i 
            ], \\[2pt]
    Z_4 = \mathrm{diag}(1, i, -1, -i).
\end{gathered}
\end{equation}
Here $X_4$ and $Z_4$ represent the generalizations of the qubit Pauli operators $X$ and $Z$, respectively, and $H_4$ denotes the ququart Hadamard.

Due to the special Hilbert space dimension $d=4=2^2$, another way to think about a ququart is as a pair of qubits. One possible encoding is given by mapping the qubit bitstrings to their corresponding decimal numbers:
\begin{equation}
\ket{q_0q_1}=
\begin{rcases}
    \begin{cases}
      \ket{00}\leftrightarrow \ket{0}\\
      \ket{01}\leftrightarrow \ket{1}\\
      \ket{10}\leftrightarrow \ket{2}\\
      \ket{11}\leftrightarrow \ket{3}
    \end{cases}  
    \hspace{-1em}
\end{rcases}=\ket*{q^{(4)}}.
\end{equation}
Therefore, two-qubit operators can be directly translated to ququart gates given their matrix representation. For example, simultaneous bit flips on $q_0$ and $q_1$ described by $X \otimes X$ can be implemented by flipping $\ket{0} \leftrightarrow \ket{3}$ and $\ket{1} \leftrightarrow \ket{2}$. Furthermore, such an encoding provides efficient qubit $C\!X$ and $SW\!AP$ gates through state flips $\ket{2} \leftrightarrow \ket{3}$ and $\ket{1} \leftrightarrow \ket{2}$, respectively. Past theoretical and experimental works have studied qubit-based computation under this scheme to find potential advantages in resource requirements and circuit speed~\cite{litteken_dancing_2023, litteken_qompress:_2023, cao_emulating_2023}.

% Therefore, two-qubit operators can be directly translated to ququart gates given their matrix representation. For example, simultaneous bit flips on $q_0$ and $q_1$ described by $X \otimes X$ can be implemented by flipping $\ket{0} \leftrightarrow \ket{3}$ and $\ket{1} \leftrightarrow \ket{2}$. While this seems more difficult than the normal qubit way, a qubit SWAP can be efficiently realized by only swapping $\ket{1} \leftrightarrow \ket{2}$. 
% Past works have theoretically studied the compression of qubits into ququarts and qubit-based computation on this higher-dimensional architecture to find potential advantages in resource requirements and circuit speed [cite Andrew].

% In this work, we will be mostly concerned with the implementation and characterization of the ququart Hadamard $H_4$ and the two-qubit Hadamard tensor product $H \otimes H$.

\subsection{Gate Decomposition}
\label{sec:decomposition}
%Efficient gate decompositions of an arbitrary unitary strongly depend on the underlying hardware. 
We use the lowest four energy levels $\{\ket{0}, \ket{1}, \ket{2}, \ket{3}\}$ of a superconducting transmon~\cite{koch_charge-insensitive_2007} to represent the ququart. The elementary gates available in the transmon are elements of $SU(2)$, meaning qubit-like rotations $R_j(\theta, \phi) = Z_j^\dagger(\phi) Y_j(\theta) Z_j(\phi)$ between two adjacent levels $\ket{j-1}$ and $\ket{j}$, with $j \in \qty{1,2,3}$. The rotations $Y_j$ around the $y$-axis are implemented using microwave pulses resonant with the $\ket{j-1} \leftrightarrow \ket{j}$ transitions, while the phase rotations $Z_j$ are realized in software (known as ``virtual $Z$'') by appropriately updating the phase of the respective carrier wave~\cite{mckay_efficient_2017} and are therefore near-perfect. The concept of virtual $Z$ gates can be generalized to the ququart case, allowing the direct implementation of relative phase rotations $Z(\vec{\phi}) = \mathrm{diag}\qty(1, e^{i \phi_1}, e^{i (\phi_1 + \phi_2)}, e^{i (\phi_1 + \phi_2 + \phi_3)})$ %of all four basis states 
by updating all three carrier waves at once, where $\vec{\phi} = (\phi_1, \phi_2, \phi_3)$ and $\phi_j$ denotes the phase shift to the $j$th microwave drive.

It has been shown that a sequence of at most six $Y_j$ rotations interleaved with phase rotations $Z$ suffices to realize any ququart unitary~\cite{dita_factorization_2003}, which is visualized in Figure~\ref{fig:decomp_pulse}(a). We modify the decomposition protocol outlined in Ref.~\cite{dita_factorization_2003} such that only one $Y_3$ rotation is required, as in our experiment the transition $\ket{2}\leftrightarrow\ket{3}$ has the smallest achievable Rabi rate. Given a ququart unitary $U$, we can numerically optimize the parameters of the sequence, where we formulate the optimization problem such that we find a sequence with minimal duration. More details can be found in Appendix~\ref{supp:gate_decomp}, where we include Table~\ref{tab:decomp} showing the obtained decompositions for $H_4$, $H \otimes H$ and a few more special gates.

\subsection{Quantum Optimal Control}
\label{sec:qoc}
Instead of constructing a target gate from high-level building blocks like pre-calibrated gates, one can take a more low-level approach and directly search for control pulses which make the quantum system undergo the desired transformation. Quantum optimal control deals with finding such optimized pulses that implement desired operations, where the optimization can be based exclusively on a simulated model (open-loop) or incorporate real hardware feedback (closed-loop). Here we use the open-source software package Juqbox~\cite{petersson_optimal_2021, petersson_discrete_2020} by LLNL as well as the proprietary tool Boulder Opal~\cite{q-ctrl_boulder_2023, ball_software_2021} from Q-CTRL to solve open-loop tasks. 

To this end we model our superconducting transmon with the Hamiltonian
\begin{align}
    \mathcal{H} = \mathcal{H}_0 + \mathcal{H}_c(t) = \sum_{n=0}^{\tilde{d}-1} \epsilon_n \op{n} -i\gamma(t)\left(a - a^\dagger\right),
\end{align}
where $\mathcal{H}_0$ and $\mathcal{H}_c$ describe the drift and the control Hamiltonian, respectively. $\gamma(t)$ is the control pulse which we write as
\begin{equation}
\begin{aligned}
    \label{eq:gamma}
    \gamma(t) &= \sum_{j=1}^3 \Gamma_j(t) \cos(\omega_j t + \varphi_j(t)) \\
    &= \Re \qty[\sum_{j=1}^3 (I_j(t) + i Q_j(t)) e^{i \omega_j t}],
\end{aligned}
\end{equation}
where for each subspace we define the quadratures $I_j(t) = \Gamma_j(t) \cos(\varphi_j(t)) = \sum_{s=1}^{N_s} \alpha_{j,s}^I B_s(t)$ and $Q_j(t) = \Gamma_j(t) \sin(\varphi_j(t))= \sum_{s=1}^{N_s} \alpha_{j,s}^Q B_s(t)$. In this work we follow the parametrization implemented in Juqbox where pulse envelopes are decomposed into $N_s$ time-local B-spline basis functions $B_s(t)$ with optimizable coefficients $\alpha^{I/Q}_{j,s}$, which allows a low-dimensional representation of long-duration pulses.

We calibrate the resonant carrier frequencies $\omega_j = \epsilon_{j} - \epsilon_{j-1}$ for the ququart subspace from Rabi and Ramsey experiments and extrapolate the energies of higher levels as $\epsilon_n = \epsilon_3 + (n-3)\omega_3 + \frac{(n-2)(n-3)}{2}(\omega_3 - \omega_2)$ for $n > 3$. This method essentially defines an anharmonicity $\xi = \omega_3 - \omega_2$ and extends the calibrated spectrum according to the anharmonic oscillator model. The Hamiltonian is truncated at $\tilde{d} = 5$ levels for simulation, which includes a guard level outside our computational subspace to capture leakage effects.

The open-loop optimal control task consists of adjusting the pulse parameters $\alpha^{I/Q}_{j,s}$ such that one obtains a pulse that minimizes the gate infidelity
\begin{align}
    1 - F_{\vec{\phi}} = 1 - \frac{1}{d^2} \abs{\Tr(U_T^\dagger Z(\vec{\phi})V_{\vec{\alpha}})}^2.
\end{align}
Here $U_T$ denotes the target unitary at time $T$ and $V_{\vec{\alpha}}$ describes the unitary the pulse with parameter vector $\vec{\alpha}$ realizes. We search for a pulse that implements the target up to a trailing phase gate $Z(\vec{\phi})$, thus we also optimize over the phase vector $\vec{\phi}$. This effectively creates a manifold of equally desirable optimization targets, some of which might be easier to realize than the original target. Since virtual phase gates come at zero cost in the experiment, this increases the freedom of the optimization and generally allows for faster pulse implementations. This trailing $Z$ gate has the same effect as $Z(\vec{\phi}_4)$ in the gate decomposition approach shown in Figure~\ref{fig:decomp_pulse}(a). Note that an explicit leading phase gate is not required as its effect is already captured by the phase relations of $I_j + i Q_j$. 

We further introduce additional cost terms to the optimization problem to ensure a narrow Fourier spectrum of the pulse $\gamma$ as well as to respect the power constraint $\sum_{j=1}^3 \frac{\abs{\Gamma_j(t)}}{r_j} \leq (1\,\mathrm{AWG})$ determined by the output limit of the arbitrary waveform generator (AWG). $r_j$ corresponds to the drive amplitude the transmon experiences when we solely drive the $j$th transition at full power, which produces a frequency-dependent response we calibrate in advance. Figure~\ref{fig:decomp_pulse}(b) shows optimized quadrature envelopes $I_j$ and $Q_j$ that modulate the carrier waves to directly realize the ququart Hadamard $H_4$ up to a final phase correction $\vec{\phi} [\mathrm{rad}] =(4.265, 4.590, 0.000)$. We similarly obtain controls for the Hadamard tensor product $H \otimes H$. For this gate the freedom of a final phase correction did not help to achieve a reduced gate duration, so we set $\vec{\phi}=\vec{0}$. More details on the drive calibration can be found in Appendix~\ref{supp:awg}, and further information on the optimal control setup as well as visualizations of pulse results is included in Appendix~\ref{supp:qoc}.

\section{Experimental Results}
We use the two-transmon processor presented in Ref.~\cite{roy2022realization, li2023hardware}, which is built from two superconducting transmons with a flux-tunable coupler, and focus solely on transmon $Q_1$. $Q_2$ remains in the ground state and can be disregarded when the coupler is not modulated and biased at the sweet spot. Details of the device are presented in Appendix \ref{supp:device}.

\begin{figure*}[t]
    \centering
    \includegraphics[width=\linewidth]{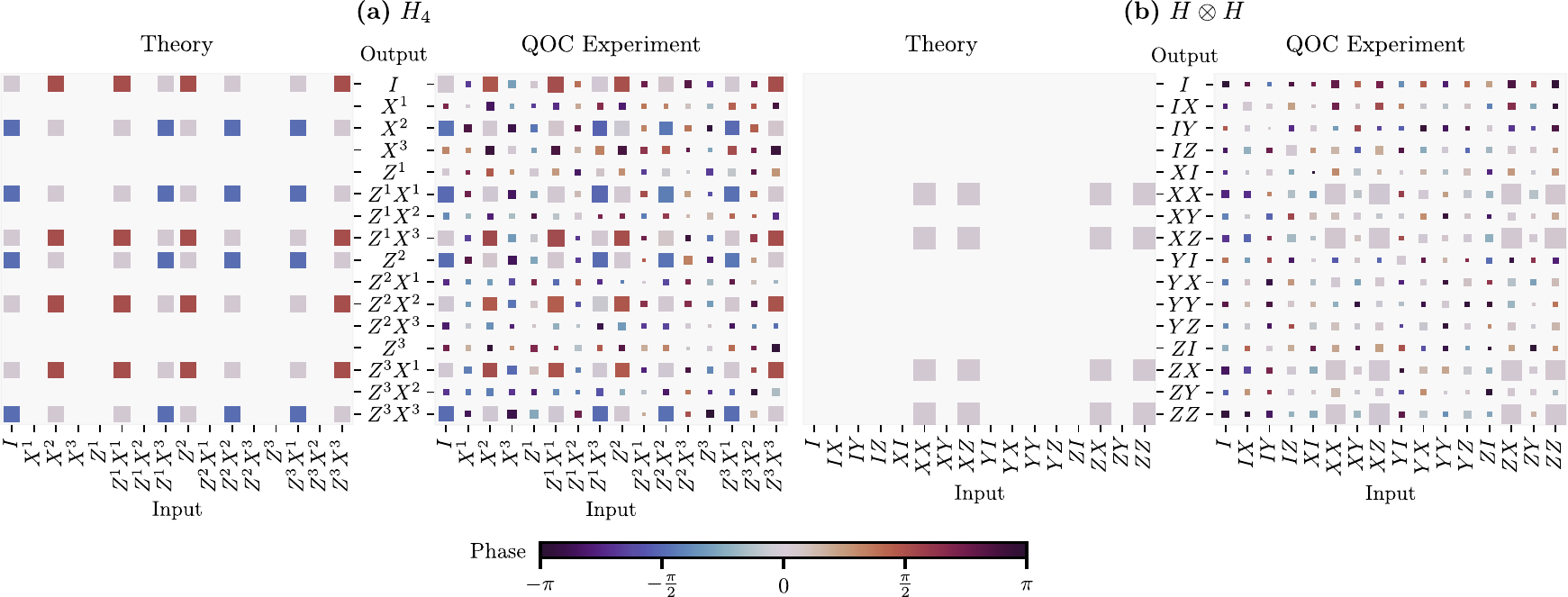}
    \caption{Process matrices for quantum optimal control implementations of the (a) ququart $H_4$ and (b) two-qubit  $H\otimes H$. The marker area encodes the magnitude of each element while the color encodes the phase. The $H_4$ process matrix is represented in the generalized Pauli basis while $H \otimes H$ is represented in the Pauli basis for two qubits.}
    \label{fig:qpt}
\end{figure*}

\subsection{Quantum Process Tomography}
We start by estimating the quality of our optimal control pulses via quantum process tomography (QPT)~\cite{chuang_prescription_1997}, which allows us to fully characterize the corresponding quantum channel $\mathcal{E}(\rho) = \sum_{j,k} \chi_{jk} B_j \rho B_k^\dagger$. We choose the operator basis $\qty{B_j}$ depending on the scenario: While for true ququart gates like the $H_4$ we use products $Z_4^m X_4^n$ of the generalized Pauli operators $X_4$ and $Z_4$, in the encoded-qubits case we follow the common choice of tensor products of qubit Pauli gates $\sigma_m \otimes \sigma_n$ and apply the encoding scheme to map them to ququart operators ($m, n \in \qty{0, 1, 2, 3}$). We extract the process matrix $\chi$ of a specific gate by applying it to 16 differently prepared states $\qty{\ket{k}, (\ket{l}+\ket{k})/\sqrt{2}, (\ket{l}-i \ket{k})/\sqrt{2}}$ ($k, l \in \qty{0,1,2,3}, k > l$) and reconstructing the resulting state using Maximum Likelihood Estimation. 

Figure~\ref{fig:qpt} shows the process matrices for both gates of interest $H_4$ and $H \otimes H$, comparing between ideal theory and implementation on our transmon. Defining the experimental process fidelity as $F = \abs{\Tr[\chi_\mathrm{ideal} \, \chi_\mathrm{expt}]}$, we obtain fidelities of $86.19 \%$ and $84.06 \%$, respectively. As a reference, we note that performing no gate (thus, characterizing the identity $\mathbb{1}_4$) leads to a process fidelity of $89.88 \%$. We trace back this loss in fidelity to measurement errors as the primary source, resulting from the challenge of implementing a four-state single-shot readout. A well-known disadvantage of quantum process tomography for benchmarking quantum gates is that it cannot separate gate errors from SPAM errors, thus yielding a lower bound on the true gate performance. We expand on the QPT configuration as well as the measurement setup in Appendix~\ref{supp:tomo_read} and include further experimental results in Table~\ref{tab:qpt_fid}.

\subsection{Randomized Benchmarking}
\label{sec:rb}
Randomized benchmarking (RB) allows extracting the average error per Clifford gate $r_\mathcal{C}$ without including SPAM errors~\cite{magesan_scalable_2011}. By averaging over randomly sampled sequences of gates from the Clifford group $\mathcal{C}$, the error channel becomes effectively depolarizing; analyzing the exponential decay over sequences of different depths $m$ allows extracting the depolarization parameter $p_\mathcal{C}$. The error per gate is then given as $r_\mathcal{C} = \frac{d-1}{d} (1-p_\mathcal{C})$ and the average gate fidelity defined as $F_\mathcal{C} = 1-r_\mathcal{C}$.

Again, we have to carefully distinguish both scenarios as the ququart Clifford group $\mathcal{C}_4$ and the two-qubit Clifford group $\mathcal{C}_2^{\otimes 2}$ differ. While $\mathcal{C}_4$ can be generated from $\qty{H_4, Z_4, S_4 = \mathrm{diag}\qty(1, \sqrt{i}, i, \sqrt{i})}$, $\mathcal{C}_2^{\otimes 2}$ is generated by the qubit gates $\qty{H, S, C\!X}$ and contains 11520 elements (up to scalars)~\cite{selinger_generators_2015}. $\mathcal{C}_4$ is not a subgroup of $\mathcal{C}_2^{\otimes 2}$ as $H_4 \notin \mathcal{C}_2^{\otimes 2}$ for instance.

\begin{figure}[htbp]
    \centering
    \includegraphics[scale=1.0]{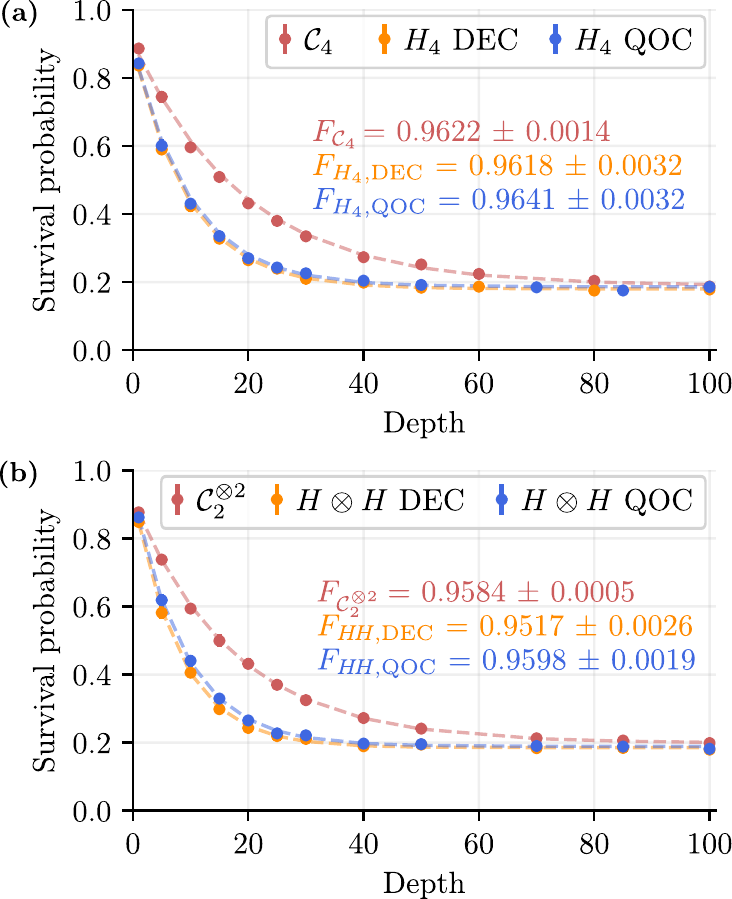}
    \caption{Experimental results for (a) ququart RB ($C_4$ Clifford group, red) and (b) two-qubit RB ($C_2^{\otimes 2}$ Clifford group, red) on a four-level transmon. Using IRB we benchmark (a) $H_4$ and (b) $H \otimes H$ in the appropriate setting, for both their gate-level decomposition ($\mathrm{DEC}$, orange) and their quantum optimal control ($\mathrm{QOC}$, blue) implementation. Markers represent the mean survival probabilities and error bars (smaller than the marker size) are the standard deviations of these means.}
    \label{fig:irb}
\end{figure}

In experiment, we perform ququart RB by sampling gates from $\mathcal{C}_4$ as well as two-qubit RB by sampling and encoding gates from $\mathcal{C}_2^{\otimes 2}$, using sequences of depth $m$ up to 100. Each sampled gate is decomposed into elementary operations as outlined in Section~\ref{sec:decomposition}. The measured survival probabilities are visualized in Figure~\ref{fig:irb} and clearly show the exponential decay. We obtain RB fidelities $F_{\mathcal{C}_4} = 96.22(14) \%$ and $F_{\mathcal{C}_2^{\otimes 2}} = 95.84(05) \%$ for the ququart RB and two-qubit RB, respectively. 
%We note that the survival probabilities do not converge to $\frac{1}{4}$ as would be expected, due to the uncertainty in our measurement.

\subsubsection*{Interleaved Randomized Benchmarking}
Interleaved RB (IRB) builds upon standard RB to characterize the error of a specific Clifford group element $G$, which is achieved by interleaving the gate of interest between the randomly sampled RB gates~\cite{magesan_efficient_2012}. Similar to standard RB, this yields the depolarization parameter $p_{\mathcal{C} + G}$. Together with the previously determined parameter $p_\mathcal{C}$, the gate fidelity can be computed as
\begin{equation}
F_G = 1 - r_G = 1 - \frac{d-1}{d} \, \qty(1-\frac{p_{\mathcal{C} + G}}{p_\mathcal{C}}).
\end{equation}

We apply IRB to characterize the ququart Hadamard $H_4$ and the Hadamard tensor product $H \otimes H$, where we interleave them between gates sampled from the appropriate Clifford groups. We do this for both the gate-level decomposition as well as the optimal control implementation of each gate and juxtapose the results in Figure \ref{fig:irb}. The open-loop optimized pulses achieve an improvement over the gate-based realizations in both cases: The $H_4$ QOC pulse is benchmarked at $96.41(32)\%$ while the decomposition achieves $96.18(32)\%$ fidelity; the $H \otimes H$ QOC pulse yields $95.98(19)\%$ fidelity compared to the decomposition with $95.17(26)\%$. This observation suggests a good agreement between the Hamiltonian model and the device. The main limitation is now caused by decoherence, dominated by the increased dephasing in the $\qty{\ket{2}, \ket{3}}$ subspace. We include more details on ququart RB as well as additional results for special gates in Appendix~\ref{supp:rb}, and present error simulations in Appendix~\ref{supp:sim}.

\section{Conclusion and Outlook}
We demonstrate the coherent control of the lowest four levels in a superconducting transmon, where we explore the implementation of both high-level optimized gate sequences and low-level optimized control pulses to realize target unitaries. While in the first approach we drive only one transition at a time, the latter approach enables manipulating all internal states at once by using open-loop quantum optimal control tools.

We study the quantum system from two angles: from the ququart perspective, as the four-dimensional generalization of a qubit, and from the perspective of two encoded qubits. We characterize gates from either perspective in QPT and RB experiments, focusing in particular on the ququart $H_4$ and the two-qubit $H \otimes H$ gate. They have similar matrix representations but belong to different Clifford groups $\mathcal{C}_4$ and $C_2^{\otimes 2}$, respectively, and we observe IRB fidelities $\geq 95\%$ for both, finding that the optimal control implementations perform better.

The fundamental challenges of increased noise on higher levels as well as the difficulty of performing a four-state readout limit the capabilities of our device. Future work should explore the application of data-driven methods like presented in Ref.~\cite{wu_data-driven_2018} to incorporate hardware feedback into the pulse calibration.

While recently presented related work also studied the two-qubit encoding scheme~\cite{cao_emulating_2023, dong_simulation_2022}, our work is, to our knowledge, the first to explore optimally controlled quantum gates on a transmon ququart and further perform ququart randomized benchmarking using the Clifford group $\mathcal{C}_4$.

\section*{Acknowledgements}
We thank Andy Goldschmidt and Jason D. Chadwick for helpful discussions on optimal control as well as N. Anders Petersson and Stefanie Günther for assistance with its implementation in Juqbox. We further thank Q-CTRL for providing us with a license for their software Boulder Opal and André Carvalho for his support in incorporating it into our work.

LMS, FTC, and JMB acknowledge funding in part by EPiQC, an NSF Expedition in Computing, under award CCF-1730449; in part by STAQ under award NSF Phy-1818914; in part by the US Department of Energy Office of Advanced Scientific Computing Research, Accelerated Research for Quantum Computing Program; and in part by the NSF Quantum Leap Challenge Institute for Hybrid Quantum Architectures and Networks (NSF Award 2016136) and in part based upon work supported by the U.S. Department of Energy, Office of Science, National Quantum Information Science Research Centers.  FTC is Chief Scientist for Quantum Software at Infleqtion and an advisor to Quantum Circuits, Inc.

ZL, TR, and DIS acknowledge support by AFOSR Grant No. FA9550-19-1-0399 and ARO Grant No. W911NF-17-S0001. The device was fabricated in the Pritzker Nanofabrication Facility at the University of Chicago, which receives support from Soft and Hybrid Nanotechnology Experimental (SHyNE) Resource (NSF ECCS-1542205), a node of the National Science Foundation’s National Nanotechnology Coordinated Infrastructure. This work also made use of the shared facilities at the University of Chicago Materials Research Science and Engineering Center.

\appendix

\section{Device parameters and fabrication}
\label{supp:device}
\begin{figure}[htbp]
    \centering
    \includegraphics[width=0.9\linewidth]{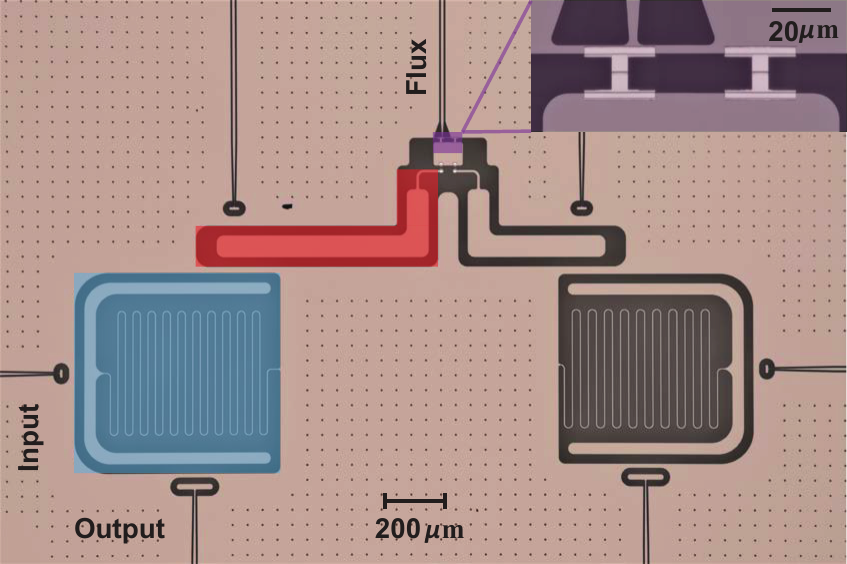}
    \caption{False-colored optical image of the device. The left transmon (red) is used as the ququart in the experiment. Single ququart rotations are sent through the input line coupled to the resonator (blue). The SQUID coupler (purple) is kept biased to the flux sweet spot throughout the experiment.}
    \centering
    \label{fig:device}
\end{figure}

\begin{figure*}[htbp]
    \centering
    \includegraphics{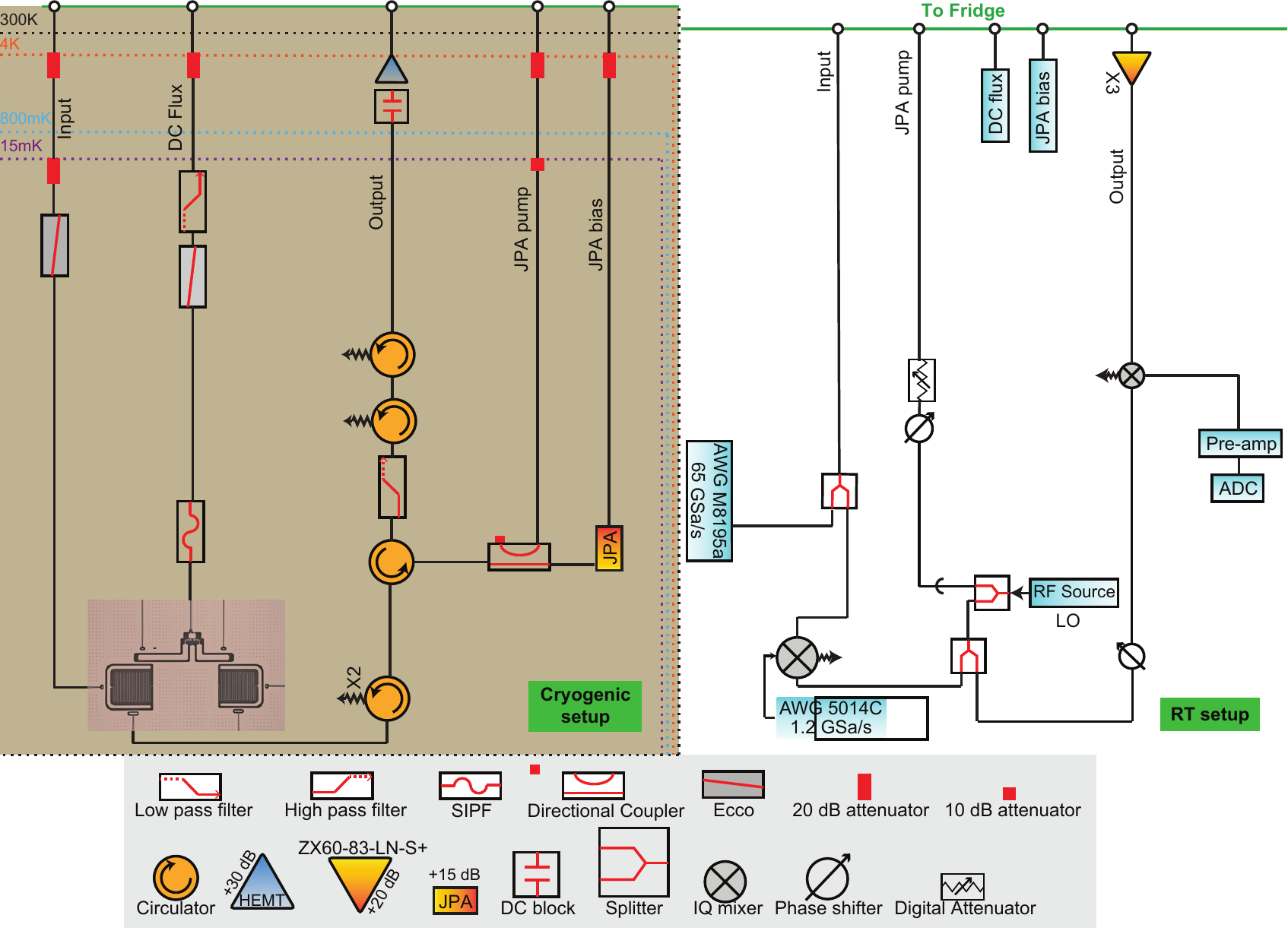}
    \caption{Measurement setup.}
    \centering
    \label{fig:setup}
\end{figure*}

The substrate used for the device fabrication is a $\SI{430}{\micro\meter}$ thick C-plane sapphire that was annealed at $1200^\circ$C for 2 hours. $\SI{200}{\nano\meter}$ thick Tantalum film was sputtered at $800^\circ$C. Large features are written with optical lithography using Heidelberg MLA 150 Aligner and wet-etched using Transene Tantalum etchant 111. Ebeam lithography with Ratih EBPG5000 Plus E-Beam writer was used to create the junction mask with a double-layer resist consisting of MMA EL11-950 and PMMA A7. The Dolan-bridge junctions were evaporated in the Plassys electron beam evaporator through double-angle evaporation. $7\times7$ mm chips are diced and lifted off. The resistance of the on-chip test junctions are measured at room temperature to help pre-selection. The selected chip was mounted on a printed circuit board, wire-bounded, and mounted inside a double-shielded $\mu$-metal can. The coherence times and frequency parameters are listed in Table~\ref{table:ququart_coherence}. Figure \ref{fig:device} shows our device's false-colored optical image. The cryogenic and room temperature measurement setup is illustrated in Figure~\ref{fig:setup}.

\begin{table}[htbp]
	\begin{tabular}{|cccc|}
	    \hline
        \hline
	     Transitions & $\ket{0}\leftrightarrow\ket{1}$ & $\ket{1}\leftrightarrow\ket{2}$ & $\ket{2}\leftrightarrow\ket{3}$ \\  \hline\hline
	    $T_{1} \ (\mu\text{s})$ & $22.62$ & $25.96$ & $10.19$ \\  \hline
	    $T_{R} \ (\mu\text{s})$ & $40.60$ & $42.32$  & $2.68$  \\  \hline
        Frequency $\omega_j/2 \pi$ (GHz) & $3.2222$ & $3.1021$ & $2.9717$ \\  \hline
        Drive response $r_j$ (MHz/AWG) & $5.5528$ & $5.6712$ & $1.6749$ \\ \hline\hline
	\end{tabular}
	%\caption{Ququart coherence and transition frequencies. Relaxation times $T_{1}$ and Ramsey dephasing times $T_{R}$ between neighboring energy levels $\ket{j-1} \leftrightarrow \ket{j}$ ($j=1,2,3$) are listed.}
    \caption{Ququart characterization. Relaxation times $T_{1}$ and Ramsey dephasing times $T_{R}$ between neighboring energy levels $\ket{j-1} \leftrightarrow \ket{j}$ ($j=1,2,3$) are listed, as well as transition frequencies $\omega_j$ and on-chip drive responses $r_j$ to unit power AWG driving.}
	\label{table:ququart_coherence}
\end{table}

\section{AWG drive calibration and native gates}
\label{supp:awg}
In experiments, all components in the measurement and control chain have their unique frequency response matrix. To map the actual ququart drive strength observed in the experiment to the drive strength specified by the input signal, we need to extract the frequency response matrix between the Arbitrary Waveform Generator (AWG) and the transmon around the frequency band used in the control pulse. In our case, we only need to consider the frequencies around the three transitions $\ket{0}\leftrightarrow\ket{1}$, $\ket{1}\leftrightarrow\ket{2}$, and $\ket{2}\leftrightarrow\ket{3}$, each within a $\SI{25}{\mega\hertz}$ frequency span.
%in the frequency domain in the optimal control. 
We further assume that under such a narrow span the frequency response matrix is constant for each transition $\ket{j-1} \leftrightarrow \ket{j}$, which allows us to directly extract the on-chip drive response factors $r_j$: $r_j$ is the experimentally measured effective drive strength at the maximum AWG output power for each transition, determined from the observed Rabi rate. For arbitrary transition rates, as they appear in the optimal control pulses, the corresponding AWG output power can be scaled accordingly. The unit of these parameters is $\qty[r_j] = \mathrm{MHz}/\mathrm{AWG}$, where we overload the abbreviation AWG to also refer to the amplitude unit of the waveform generator. We include the calibrated $r_j$ values in Table~\ref{table:ququart_coherence}. 

For all physical rotations in the decomposed gate sequence, the pulse shape is a Gaussian flat-top pulse with a $2\sigma = \SI{5}{\nano\second}$ ramp and a $2\sigma = \SI{5}{\nano\second}$ tail, where $\sigma$ is the standard deviation of the Gaussian edge. In order to implement an arbitrary rotation angle $\theta$ in the $j$th subspace, we tune the duration of the constant part $\tau_j$ of the flat-top section while keeping the power unchanged. The overall gate duration $T_j(\theta) = \tau_j(\theta) + 4 \sigma$ is then given by:
\begin{equation}
\label{eq:decomp_time}
\begin{aligned}
    T_j(\theta) &= \frac{\theta}{2\pi \sqrt{j}\, r_j \cdot 1~\mathrm{AWG}} - \sqrt{2\pi} \mathrm{erf}\qty(\sqrt{2})\sigma + 4\sigma \\
    &\approx \frac{\theta}{2\pi \sqrt{j}\, r_j \cdot 1~\mathrm{AWG}} + \SI{4}{\nano\second}.
\end{aligned}
\end{equation}

\section{Numerical gate decomposition}
\label{supp:gate_decomp}
We elaborate on finding the parameters of the gate decomposition outlined in Section \ref{sec:decomposition} given a target unitary $U$. Our native operation set consists of rotations $R_j(\theta, \phi) = Z_j^\dagger(\phi) Y_j(\theta) Z_j(\phi)$ in the subspaces $\qty{\ket{j-1}, \ket{j}}$, $j=\{1,2,3\}$, which explicitly read:
\begin{subequations}
\label{eq:single_qutrit}
\begin{align}
    R_{1}\left(\phi,\theta\right) &= 
    \begin{bmatrix}
               \cos{\frac{\theta}{2}} & -e^{i\phi}\sin{\frac{\theta}{2}} & 0 & 0 \\
            e^{-i\phi}\sin{\frac{\theta}{2}} & \cos{\frac{\theta}{2}} & 0 & 0 \\
           0 & 0 & 1 & 0 \\
           0 & 0 & 0 & 1
    \end{bmatrix},
    \\
    R_{2}\left(\phi,\theta\right) &= 
    \begin{bmatrix}
               1 & 0 & 0 & 0 \\
           0 & \cos{\frac{\theta}{2}} & -e^{i\phi}\sin{\frac{\theta}{2}} & 0 \\
           0 & e^{-i\phi}\sin{\frac{\theta}{2}} & \cos{\frac{\theta}{2}} & 0 \\
           0 & 0 & 0 & 1
    \end{bmatrix},
    \\
    R_{3}\left(\phi,\theta\right) &= 
    \begin{bmatrix}
               1 & 0 & 0 & 0 \\
               0 & 1 & 0 & 0 \\
           0 & 0 & \cos{\frac{\theta}{2}} & -e^{i\phi}\sin{\frac{\theta}{2}} \\
           0 & 0 & e^{-i\phi}\sin{\frac{\theta}{2}} & \cos{\frac{\theta}{2}}           
    \end{bmatrix}.
\end{align}
\end{subequations}

As derived in Ref.~\cite{dita_factorization_2003} and shown in Figure~\ref{fig:decomp_pulse}, every single-ququart gate $U$ can be decomposed in the following way:
\begin{equation}
\label{eq:decomposition}
\begin{aligned}
    U &= Z(\vec{\phi}_4)Y_1(\theta_6) Y_2(\theta_5) Y_3(\theta_4) Z(\vec{\phi}_3) \\
    & \quad \times Y_1(\theta_3) Y_2(\theta_2) Z(\vec{\phi}_2) Y_1(\theta_1) Z(\vec{\phi}_1).
\end{aligned}
\end{equation}
Here $Z(\vec{\phi}_i)$ is a virtual ququart phase rotation which implements all subspace phase shifts $\vec{\phi}_i = (\phi_{i,1}, \phi_{i,2}, \phi_{i,3})$ at once. This decomposition holds up to a global phase, which is inconsequential in the experiment. While at first, it looks like $4 \times 3 \text{ phases} + 6 \text{ angles} = 18$ real degrees of freedom, which contradicts the 15  real degrees of freedom for an element of $SU(4)$, it is actually the case that one can fix $\phi_{1,2} = \phi_{1,3} = \phi_{2,3} = 0$~\cite{dita_factorization_2003}, leaving the right number of parameters. We changed the indexing of the subspaces compared to Ref.~\cite{dita_factorization_2003} such that the decomposition uses the fewest gates in the third subspace. This was motivated by the fact that we experimentally observed a significantly reduced achievable Rabi rate in this subspace as presented in Appendix~\ref{supp:awg}, which would likely cause much slower gate implementations when using the original decomposition under the same experimental constraints. We emphasize that Equation~\eqref{eq:decomposition} represents the most general decomposition and in special cases not all gates might be needed. For instance, we have $X_4 = Z(\pi,0,0) Y_1(\pi) Y_2(\pi) Y_3(\pi)$ but $H_4$ requires all gates as shown in Table~\ref{tab:decomp}.

Given a target unitary $U$, we determine the decomposition parameters $\theta_i \in [0, \pi]$ and $\vec{\phi}_i \in [0, 2\pi)^{\otimes 3}$ via numerical optimization, where we primarily minimize the gate infidelity
\begin{equation}    
    1-F = 1 - \frac{1}{d^2}\abs{\Tr(U^\dagger V)}^2
    \label{eq: infidelity}
\end{equation}
between the target and the decomposition $V = V(\theta_i, \vec{\phi}_i)$. We speak of a \textit{feasible decomposition} if $1-F \leq 10^{-8}$. 

For a specific target, there may be many feasible decompositions, however, not all of them need to be reasonable with respect to experimental implementation. Consider the identity operation: $\mathbb{1} = Y_1(\pi)Y_1(\pi)$ would be compatible with the general formulation \eqref{eq:decomposition} but unnecessarily expensive to implement. Therefore we also want to minimize the estimated overall gate duration
\begin{align}
    \label{eq:decomp_time_est}
    T^*(\theta_i) = \sum_{i=1}^6 \frac{\theta_i}{2\pi \sqrt{j} \, r_{s(i)} \cdot 1\,\mathrm{AWG}}
\end{align}
which is determined by the rotation angles. For simplicity, we assume rectangular pulse shapes in this model and consider driving at the maximum AWG output like in the experiment. The function $s$ maps the angle index $i$ to the corresponding subspace index $j$. 
%We note that this linear duration scaling with rotation angle rather serves as an estimate since our native pulses are not perfectly rectangular but have Gaussian edges. 
However, this approach is not ideal because it could lead to the unnecessary splitting of operations. For instance, if $U = Y_1(\pi)$, feasible decompositions would be $U=Y_1(\pi)$ (ideal) or $U=Y_1(\frac{\pi}{3})Y_1(\frac{\pi}{3})Y_1(\frac{\pi}{3})$ (not ideal). Therefore we actually use a different penalty of the form
\begin{align}
    \tilde{T}({\theta_i}) = \sum_{i=1}^6 \sqrt{\frac{\theta_i}{2\pi \sqrt{j} \, r_{s(i)} \cdot 1\,\mathrm{AWG}}}.
\end{align}
This makes use of the fact that, for $a,b \geq 0$, $\sqrt{a+b} \leq \sqrt{a} + \sqrt{b}$, with equality if and only if $a$ or $b$ equals zero. Thus splitting of native rotations is discouraged. Other than that $\tilde{T}$ is still monotonically increasing in each $\theta_i$ and minimizing $\tilde{T}$ also leads to reducing $T^*$. 

Additionally, it may be the case that (infinitely) many sets of phase angles $\vec{\phi}_i$ lead to feasible decompositions, even for optimal rotation angles. In order to reduce this degeneracy and furthermore get a better intuition of the action of the gate sequence, we steer the optimization towards phase angles that are multiples of $\frac{\pi}{2}$. In conjunction with $Y_j$ gates, these special phase shifts effectively lead to rotations about the $\pm x$-axis and $\pm y$-axis in the $j$th subspace's Bloch sphere. We achieve this by introducing another cost term
\begin{align}
    P(\vec{\phi}_i) = \frac{1}{12} \sum_{i=1}^4\sum_{j=1}^3 \sin^2(2\phi_{i,j}).
\end{align}
Overall, for a given target $U$ we solve the minimization problem
\begin{align}
    \hat{\theta}_i, \hat{\vec{\phi}}_i = \arg\min_{\theta_i, \vec{\phi}_i} I + c_{\tilde{T}} \tilde{T} + c_P P
\end{align}
to find a feasible decomposition. We optimize a batch of ten initial guesses for the optimization variables and check if the best-obtained infidelity is below the desired threshold; if not, we keep optimizing batches until we do. For a suitable choice of penalty coefficients $c_{\tilde{T}}$ and $c_P$ this method is guaranteed to succeed eventually since any unitary can be realized with decomposition \eqref{eq:decomposition}. We set $c_{\tilde{T}} = 0.2$ to strongly encourage fast sequences and $c_P = 0.05$ to gently push the optimizer towards preferred phase values, and find this setup to work consistently well in practice.

We show the decomposition for the ququart Hadamard $H_4$ as well as several gates acting on two qubits in Table \ref{tab:decomp} and further list the execution time computed by summing up the individual pulse times according to Equation~\eqref{eq:decomp_time}, rounded to full ns. The numerically obtained parameter values suggest that they approximate multiples of $\frac{\pi}{2}$ or more complicated analytical expressions: For example, for the gate $H_4$ we find $\theta_1 \approx 2.09440$, which has the property $e^{i\theta_1/2} \approx \frac{1+\sqrt{3}i}{2}$. This analytical nature is reasonable given the structure of the gates we consider to factor unitaries, which motivates studying qudit gate decompositions further and finding better ways to obtain these expressions directly.

\addtolength{\tabcolsep}{3pt}
\begin{table}[htbp]
    \centering
    \begin{tabular}{|c cccccc c|}
        \hline \hline
        Gate   & $\theta_{1,2,3}$   & $\theta_{4,5,6}$   & $\vec{\phi}_1$   & $\vec{\phi}_2$   & $\vec{\phi}_3$   & $\vec{\phi}_4$   & $T [\mathrm{ns}]$ \\ \hline \hline
                     & $\frac{\pi}{2}$    & $\tilde{\theta}_1$ & $\frac{3}{2}\pi$ & $\tilde{\phi}_1$ & $\tilde{\phi}_1$ & $\frac{\pi}{2}$  &     \\
        $H_4$        & $\tilde{\theta}_2$ & $\tilde{\theta}_2$ & 0                & $\frac{3}{2}\pi$ & $\tilde{\phi}_2$ & $\frac{\pi}{2}$  & 343 \\
                     & $\tilde{\theta}_3$ & $\frac{\pi}{2}$    & 0                & 0                & $\frac{3}{2}\pi$ & $\frac{\pi}{2}$  &     \\ \hline
                     & $\frac{\pi}{2}$    & $\tilde{\theta}_1$ & $\pi$            & $\pi$            & $\pi$            & 0                &     \\
       $H\otimes H$  & $\tilde{\theta}_2$ & $\tilde{\theta}_2$ & 0                & 0                & $\pi$            & $\pi$            & 365 \\
                     & $\tilde{\theta}_1$ & $\frac{\pi}{2}$    & 0                & 0                & $\pi$            & 0                &     \\ \hline
                     & $\frac{\pi}{2}$    & $\frac{\pi}{2}$    & $\pi$            & 0                & $\frac{\pi}{2}$  & $\frac{3}{2}\pi$ &     \\
$\mathbb{1}\otimes H$& 0                  & 0                  & 0                & 0                & $\frac{\pi}{2}$  & $\frac{\pi}{2}$  & 139 \\
                     & 0                  & 0                  & 0                & 0                & $\pi$            & 0                &     \\ \hline
                     & 0                  & $\frac{\pi}{2}$    & 0                & $\pi$            & 0                & $\pi$            &     \\
$H\otimes \mathbb{1}$& $\pi$              & $\pi$              & 0                & $\pi$            & $\pi$            & $\pi$            & 272 \\
                     & $\frac{\pi}{2}$    & 0                  & 0                & 0                & 0                & 0                &     \\ \hline 
                     & 0                  & $\pi$              & 0                & 0                & 0                & 0                &     \\
$C\!X_{c=q_0}^{t=q_1}$&0                  & 0                  & 0                & 0                & 0                & 0                & 176 \\
                     & 0                  & 0                  & 0                & 0                & $\pi$            & 0                &     \\ \hline
                     & 0                  & $\pi$              & 0                & $\pi$            & $\pi$            & $\pi$            &     \\
$C\!X_{c=q_1}^{t=q_0}$& $\pi$             & $\pi$              & 0                & $\frac{3}{2}\pi$ & $\pi$            & $\frac{\pi}{2}$  & 309 \\
                     & 0                  & 0                  & 0                & 0                & $\frac{\pi}{2}$  & $\frac{3}{2}\pi$ &      \\ \hline \hline
    \end{tabular}
    \addtolength{\tabcolsep}{-3pt}
    
    \bigskip

    \addtolength{\tabcolsep}{2pt}
    \begin{tabular}{|cccccc|}
        \hline \hline
        & $\tilde{\theta}_1$ & $\tilde{\theta}_2$ & $\tilde{\theta}_3$ & $\tilde{\phi}_1$ & $\tilde{\phi}_2$ \\ \hline \hline
        value & 2.09440 & 1.91063 & 1.31812 & 5.17604 & 5.03414 \\
        $e^{i \tilde{\theta}/2}$ & $\frac{1+\sqrt{3}i}{2}$ & $\frac{1+\sqrt{2}i}{\sqrt{3}}$ & $\frac{\sqrt{5}+\sqrt{3}i}{\sqrt{8}}$ & $-$ & $-$ \\
        $e^{i \tilde{\phi}}$ & $-$ & $-$ & $-$ & $\frac{\sqrt{2}-\sqrt{8}i}{\sqrt{10}}$ & $\frac{1-3i}{\sqrt{10}}$ \\ \hline \hline
    \end{tabular}
    \addtolength{\tabcolsep}{-2pt}
    \caption{Numerically obtained decomposition parameters for several gates according to Equation \eqref{eq:decomposition}, together with sequence execution durations $T$ computed using Equation \eqref{eq:decomp_time}. All angles are measured in radians. We recognize decimal approximations of parameters as multiples of $\frac{\pi}{2}$ or special analytical expressions which we show in the lower table.}
    \label{tab:decomp}
\end{table}

% \begin{table}[]
%     \centering
%     \begin{tabular}{c|c}
%         $Z$ & $(\frac{3\pi}{2}, 0, 0)$ \\
%         $Y_0$ & $\frac{\pi}{2}$ \\
%         $Z$ & $(\lambda_1, \frac{3\pi}{2}, 0)$ \\
%         $Y_1$ & $\xi_1$ \\
%         $Y_0$ & $\xi_2$ \\
%         $Z$ & $(\lambda_1, \lambda_2, \frac{3\pi}{2})$ \\
%         $Y_2$ & $\xi_3$ \\
%         $Y_1$ & $\xi_1$ \\
%         $Y_0$ & $\frac{\pi}{\2}$ \\
%         $Z$ & $(\frac{\pi}{2}, \frac{\pi}{2}, \frac{\pi}{2})$
%     \end{tabular}
%     \caption{$\cos(\frac{\xi_1}{2}) = \sqrt{3}^{-1}$, $\cos(\frac{\xi_2}{2}) = \sqrt{\frac{5}{8}}$, $\cos(\frac{\xi_1}{2}) = \sqrt{4}^{-1}$, $\cos(\lambda_1) = \sqrt{3}^{-1}$, $\cos(\frac{\xi_1}{2}) = \sqrt{3}^{-1}$}
%     \label{tab:my_label}
% \end{table}

\section{Optimal control}
\label{supp:qoc}
Our native as well as optimal-control pulses implement a desired gate $U_T$ (duration $T$) in the interaction frame defined by $\ket{\psi_\mathrm{rot}(t)} = e^{i \mathcal{H}_0 t} \ket{\psi_\mathrm{lab}(t)}$ with Hamiltonian
\begin{equation*}
    \mathcal{H}_\mathrm{int} = \sum_{j=1}^3 \frac{\sqrt{j}}{2} (I_j(t) Y_j + Q_j(t) X_j),
\end{equation*}
where $I_j$ and $Q_j$ are the quadrature signals for the $j$th subspace. In the optimal control setting they are built from time-local B-spline basis functions like defined in Section~\ref{sec:qoc} and used in Juqbox~\cite{petersson_optimal_2021}:
\begin{align}
    I_j(t) &= \sum_{s=1}^{N_s} \alpha_{j,s}^I B_s(t), & Q_j(t) &= \sum_{s=1}^{N_s} \alpha_{j,s}^Q B_s(t).
\end{align}
However, computation in this frame cannot take the leakage into higher energy levels into account, therefore we consider a different rotating frame transformation $\ket{\psi_\mathrm{rot}(t)} = e^{i \omega_\mathrm{rot} t \hat{n}} \ket{\psi_\mathrm{lab}(t)}$ to solve the open-loop optimal control problem instead. This yields the Hamiltonian
\begin{equation}
\begin{aligned}
    \mathcal{H}_\mathrm{rot}(t) = &\sum_{n=0}^{\tilde{d}-1} \qty(\epsilon_n - n\omega_\mathrm{rot}) \op{n} \\
    &+ \frac{1}{2} \qty(I(t) (-i) \qty(a - a^\dag) + Q(t) \qty(a + a^\dag))
\end{aligned}
\end{equation}
with $I(t) + iQ(t) = \sum_{j=1}^3 \qty(I_j(t) + iQ_j(t)) e^{i\qty(\omega_j - \omega_\mathrm{rot})t}$. The Hamiltonian is truncated at $\tilde{d}=5$ levels to account for leakage and we find this to be sufficient in simulation (see Appendix~\ref{supp:sim}). Accordingly, the target gate has to be transformed to $U'_T = W_{\mathrm{trans}, T} U_T$, where $W_{\mathrm{trans}, T} = e^{i \omega_\mathrm{rot} T \hat{n}} e^{-i \mathcal{H}_0 T}$. The infidelity cost term becomes
\begin{align}
    1 - F_{\vec{\phi}} = 1 - \frac{1}{d^2} \abs{\Tr(U_T'^\dagger Z(\vec{\phi})V_{\vec{\alpha}})}^2.
\end{align}
with the phase corrections $\vec{\phi}$ and pulse parameters $\vec{\alpha} = \qty(\alpha_{j,s}^I, \alpha_{j,s}^Q)_{j,s}$ that are optimized over. 

We choose $\omega_\mathrm{rot} = \omega_1$ to reduce the oscillating behavior of the functions $I$ and $Q$, which allows the choice of larger time steps in the numerical solution of the dynamics. Typically we set the time step size $\Delta t = 0.03\dots 0.3 \, \si{\nano\second}$. In most optimal control scenarios the gate time $T$ needs to be fixed a priori and infidelity convergence is an indicator if the chosen duration was sufficient to realize the gate. For the pulse parametrization at hand, we choose the number of B-splines $N_s$ proportional to the gate time, $N_s = \left\lfloor\frac{T}{\SI{10}{\nano\second}}\right\rfloor$, to ensure similar B-spline densities for pulses of different durations. 

\begin{figure}[htbp]
    \centering
    \includegraphics[scale=0.8]{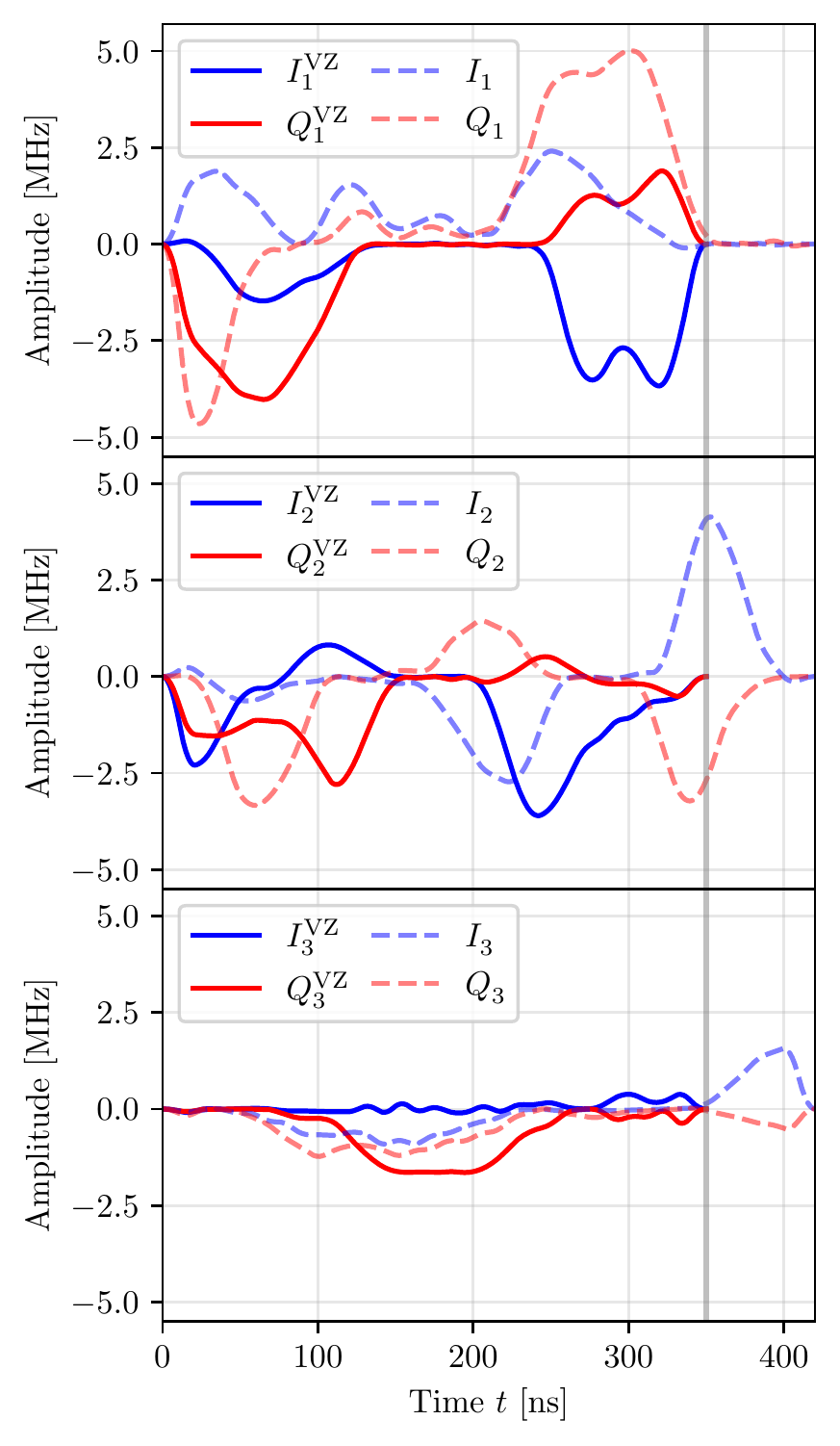}
    \caption{Optimal control pulses for ququart Hadamard $H_4$, with final VZ correction (solid lines, $T=\SI{350}{\nano\second}$) and without final VZ correction (dashed lines, $T=\SI{420}{\nano\second}$). The grey line marks the duration of the former.}
    \label{fig:iq_h4}
\end{figure}

We try to find optimal control pulses of short durations as decoherence limits the capabilities of our device, especially due to increased dephasing in the $\qty{\ket{2}, \ket{3}}$ subspace (see Table~\ref{table:ququart_coherence}). In this work, we perform the duration optimization by hand, which is achieved by reducing the gate time further and further while making sure the numerical infidelity stays below the threshold $1 - F_{\vec{\phi}} \leq 10^{-4}$. We note that this is a rather simple method that could be improved. One alternative approach would be to make the duration $T$ an optimization variable, rescale all time variables with $\frac{1}{T}$ and all frequency variables with $T$, respectively, and then solve the optimal control task in the time interval $[0, 1]$. However, we observe worse convergence and higher obtained gate durations in this case. Despite its simpler nature, our method is sufficient for our purposes and manages to show the impact of performing final phase corrections using a VZ on the gate time $T$. Figure~\ref{fig:iq_h4} shows $H_4$ pulse implementations both with (solid lines) and without a final VZ (dashed lines), where the difference in duration is highlighted by the solid grey line. Interestingly we find that for the $H \otimes H$ tensor product optimizing with a final VZ does not lead to a reduced gate duration, therefore we choose the optimization result without a VZ in our experiments.

In the following, we provide details on the additionally introduced cost terms like mentioned in Section~\ref{sec:qoc}. One important hardware constraint the optimized control pulse needs to obey is the power limit of the arbitrary waveform generator (AWG), which imposes $\abs{\gamma(t)} \leq 1\,\mathrm{AWG}$. Here we overload the abbreviation $\mathrm{AWG}$ to use it as the corresponding power unit and further reiterate that the conversion from the drive strength on-chip (in MHz) to AWG power is frequency-dependent. With Equation \eqref{eq:gamma} this constraint can be ensured by imposing
\begin{subequations}
\begin{align}
    |\gamma(t)| \leq \sum_{j=1}^3 \frac{|\Gamma_j(t)|}{r_j} \equiv \tilde{\Gamma}(t) \overset{!}{\leq} 1\,\mathrm{AWG} \\
    \label{eq: alpha_constraint}
    \Rightarrow \sum_{j=1}^{3} \frac{1}{r_j} \sqrt{(\alpha_{j,s}^I)^2 + (\alpha_{j,s}^Q)^2} \leq 1\,\mathrm{AWG}.
\end{align}
\end{subequations}
At this time Juqbox does not support constraints that include multiple optimization variables. Without any additional knowledge of the structure of the target gate, the most general approach is assigning one third of the power budget to each of the three carrier waves, which can be done by enforcing $\abs{\alpha_{j,s}^I}, \abs{\alpha_{j,s}^Q} \leq \frac{r_j}{3 \sqrt{2}}$. However, this may drastically reduce the potential speed of the gate implementation as no transition is allowed to be driven at more than a third of the maximum power.

We transition to using Boulder Opal~\cite{q-ctrl_boulder_2023} while sticking to the Juqbox pulse parametrization to overcome this limitation. While Boulder Opal currently cannot directly implement constraints of the form \eqref{eq: alpha_constraint} either, it allows for a more flexible formulation of the quantum control task, construction of signals, and definition of cost terms. Let us consider the functions
\begin{align}
    f(x) &= (1-g(x-a))x + g(x-a)a, \\
    g(x) &= 1-\frac{\tanh(-\frac{1000x}{a})+1}{2},
\end{align}
where $g$ serves as a differentiable version of the Heaviside step function $\Theta(x)$ that can be implemented in Boulder Opal's framework. $f$ can be seen as a suppression function with threshold $a$, which means $f(x) \approx x$ for $x < a$ and $f(x) \approx a$ for $x \geq a$. Using this we can construct a suppressed signal
\begin{align}
    \tilde{\Gamma}^\text{sup}(t) = f\qty(\tilde{\Gamma}(t))
\end{align}
and a penalty term
\begin{align}
    P_\text{amp} = \int_0^T \abs{\tilde{\Gamma}(t) - \tilde{\Gamma}^\text{sup}(t)} \text{d}t,
\end{align}
which captures how much the original signal $\tilde{\Gamma}$ violates the power constraint. We set the threshold amplitude to $a=0.95\,\mathrm{AWG}$ to allow the optimizer some slack to the critical value of $1\,\mathrm{AWG}$.

The second cost term addresses the assumption of a constant frequency response matrix in the range $\qty[\SI{-25}{\mega\hertz}, \SI{25}{\mega\hertz}]$ around the three transition frequencies, which we discuss in Appendix~\ref{supp:awg}. In order to be consistent with this assumption in the optimal control pulse design, we introduce a filter penalty to steer the optimization towards pulses with narrow peaks in the Fourier spectrum:
\begin{align}
    P_\text{fil} = \sum_{j=1}^3  \int_0^T \abs{\Gamma_j(t) - \Gamma_j^\text{fil}(t)} \text{d}t.
\end{align}
where $\Gamma_j^\text{fil}$ is the convolution of $\Gamma_j$ with a $\operatorname{sinc}$-kernel to limit the bandwidth around each transition to $\qty[\SI{-25}{\mega\hertz}, \SI{25}{\mega\hertz}]$. The computation of the convolution is a built-in feature in Boulder Opal. 

\begin{figure}[htbp]
    \centering
    \includegraphics[scale=0.8]{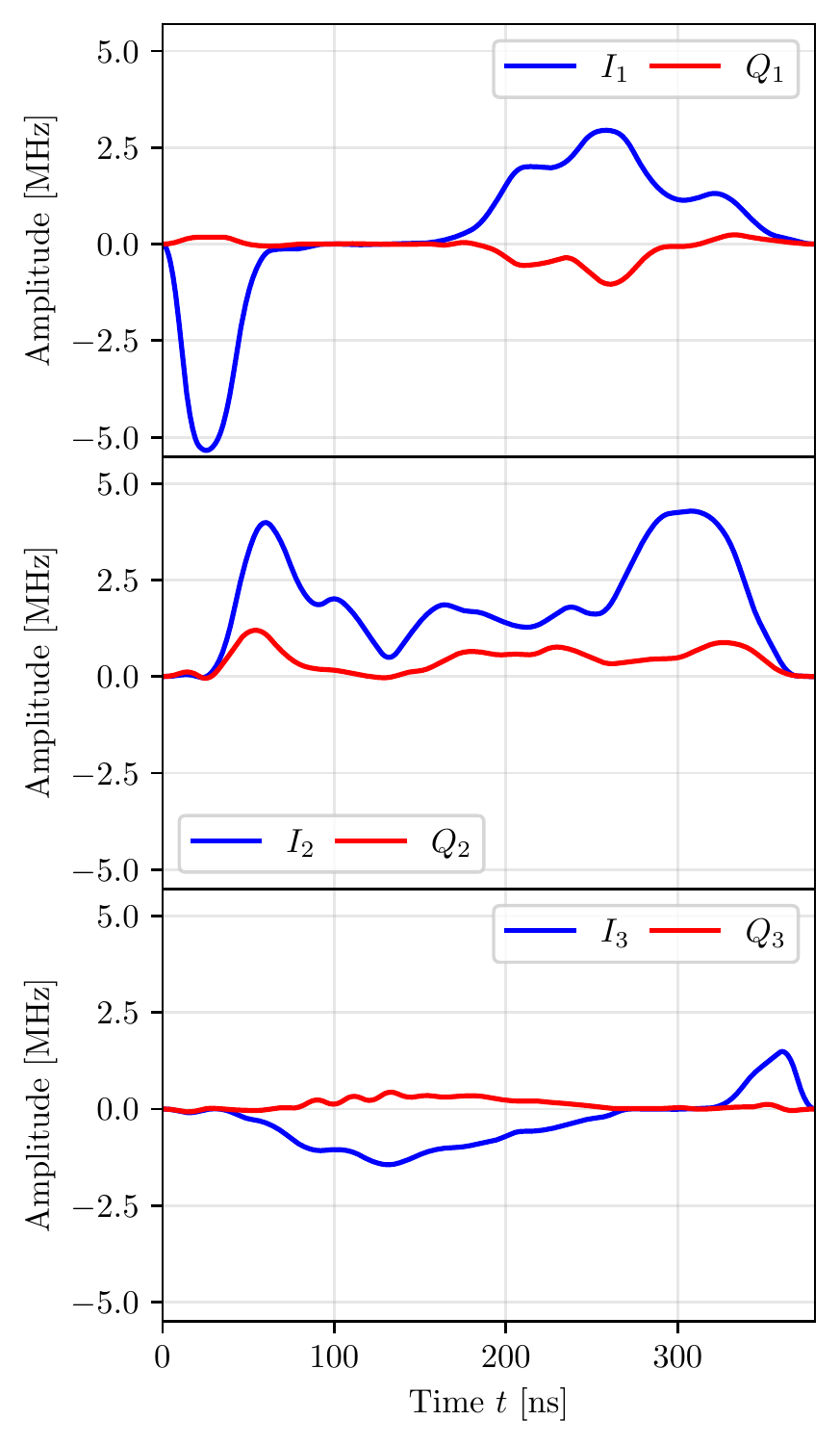}
    \caption{Optimal control pulses for the tensor product $H \otimes H$ without a final VZ correction, $T=\SI{380}{\nano\second}$. Allowing a correction did not lead to a significantly reduced gate duration.}
    \label{fig:iq_hh}
\end{figure}

The full optimal control problem then reads
\begin{align}
    \hat{\alpha}_{j,s}, \hat{\vec{\phi}} = \arg \min_{\alpha_{j,s}, \vec{\phi}} I_{\vec{\phi}} + c_\text{amp} P_\text{amp} + c_\text{fil} P_\text{fil},
\end{align}
where we empirically choose $c_\text{amp}=c_\text{fil}=\frac{1}{2}$. We show the obtained pulse implementations for our gates of interest $H_4$ and $H \otimes H$ in Figures~\ref{fig:iq_h4} and \ref{fig:iq_hh}, respectively.

% We use a modified version of the infidelity $I_{\vec{\phi}}$ which measures the distance to the target gate up to a final virtual phase gate:
% \begin{align}
%     I_{\vec{\phi}} = 1 - \frac{1}{16} |\Tr(U^\dagger Z(\vec{\phi})V)|^2.
% \end{align}
% This effectively creates a manifold of optimization targets, where some might be easier to realize than the original target. Since virtual phase gates come at zero cost in experiments, this increases the freedom of the optimization and generally allows for faster pulse implementations. This trailing $Z$ gate has the same effect as $Z(\vec{\phi}_4)$ in the decomposition formula \eqref{eq:decomposition}.

% [Show example pulse, w/ and w/o VZ]

\section{Tomography and readout histogram}
\label{supp:tomo_read}
To perform single-ququart quantum state tomography (QST), we apply one of 16 different post-rotation gates prior to the single-shot readouts. This set $\mathcal{T}$ is shown in Table \ref{tab:pre-post}, where $\mathbb{1}$ is the identity gate and the rotations $R_{1}$, $R_{2}$, and $R_{3}$ are defined in Appendix~\ref{supp:gate_decomp}.
% $S=\{\}$
% we apply 16 post rotations $S_{j}$ from the tomography rotation set $S$: $S=\{I, R_{1}\left(0,\frac{\pi}{2}\right), R_{1}\left(\frac{\pi}{2},\frac{\pi}{2}\right),R_{1}\left(0,\pi\right),R_{2}\left(0,\frac{\pi}{2}\right),$\\
% $R_{2}\left(\frac{\pi}{2},\frac{\pi}{2}\right),R_{2}\left(0,\frac{\pi}{2}\right)R_{1}\left(0,\pi\right),R_{2}\left(\frac{\pi}{2},\frac{\pi}{2}\right)R_{1}\left(0,\pi\right),$\\
% $R_{2}\left(0,\pi\right)R_{1}\left(0,\pi\right), R_{3}\left(0,\frac{\pi}{2}\right), R_{3}\left(\frac{\pi}{2},\frac{\pi}{2}\right),$\\ $R_{3}\left(0,\pi\right)R_{2}\left(0,\pi\right)R_{1}\left(0,\pi\right),$\\
% $R_{3}\left(0,\frac{\pi}{2}\right)R_{2}\left(0,\pi\right)R_{1}\left(0,\pi\right),$\\
% $R_{3}\left(\frac{\pi}{2},\frac{\pi}{2}\right)R_{2}\left(0,\pi\right)R_{1}\left(0,\pi\right),$\\
% $R_{3}\left(0,\frac{\pi}{2}\right)R_{2}\left(0,\pi\right), R_{3}\left(\frac{\pi}{2},\frac{\pi}{2}\right)R_{2}\left(0,\pi\right)\}$. 
\begingroup
\setlength{\tabcolsep}{6pt}
\begin{table}[htbp]
    \centering
    \begin{tabular}{|cc|}
        \hline \hline
        Pre-rotation set $\mathcal{S}$ & Post-rotation set $\mathcal{T}$ \\
        \hline \hline
        $\mathbb{1}_4$ & $\mathbb{1}_4$ \\
        $R_{1}\left(0,\frac{\pi}{2}\right)$ & $R_{1}\left(0,\frac{\pi}{2}\right)$ \\
        $R_{1}\left(\frac{\pi}{2},\frac{\pi}{2}\right)$ & $R_{1}\left(\frac{\pi}{2},\frac{\pi}{2}\right)$ \\
        $R_{1}\left(0,\pi\right)$ & $R_{1}\left(0,\pi\right)$ \\
        $R_{2}\left(0,\frac{\pi}{2}\right)$ & $R_{2}\left(0,\frac{\pi}{2}\right)$ \\
        $R_{2}\left(\frac{\pi}{2},\frac{\pi}{2}\right)$ & $R_{2}\left(\frac{\pi}{2},\frac{\pi}{2}\right)$ \\
        $R_{2}\left(0,\pi\right)R_{1}\left(0,\frac{\pi}{2}\right)$ & $R_{2}\left(0,\frac{\pi}{2}\right)R_{1}\left(0,\pi\right)$ \\
        $R_{2}\left(0,\pi\right)R_{1}\left(\frac{\pi}{2},\frac{\pi}{2}\right)$ & $R_{2}\left(\frac{\pi}{2},\frac{\pi}{2}\right)R_{1}\left(0,\pi\right)$ \\
        $R_{2}\left(0,\pi\right)R_{1}\left(0,\pi\right)$ & $R_{2}\left(0,\pi\right)R_{1}\left(0,\pi\right)$ \\
        $R_{3}\left(0,\frac{\pi}{2}\right)$ & $R_{3}\left(0,\frac{\pi}{2}\right)$ \\
        $R_{3}\left(\frac{\pi}{2},\frac{\pi}{2}\right)$ & $R_{3}\left(\frac{\pi}{2},\frac{\pi}{2}\right)$ \\ 
        $R_{3}\left(0,\pi\right)R_{2}\left(0,\frac{\pi}{2}\right)$ & $R_{3}\left(0,\frac{\pi}{2}\right)R_{2}\left(0,\pi\right)$ \\
        $R_{3}\left(0,\pi\right)R_{2}\left(\frac{\pi}{2},\frac{\pi}{2}\right)$ & $R_{3}\left(\frac{\pi}{2},\frac{\pi}{2}\right)R_{2}\left(0,\pi\right)$ \\
        $R_{3}\left(0,\pi\right)R_{2}\left(0,\pi\right)R_{1}\left(0,\frac{\pi}{2}\right)$ & $R_{3}\left(0,\frac{\pi}{2}\right)R_{2}\left(0,\pi\right)R_{1}\left(0,\pi\right)$ \\
        $R_{3}\left(0,\pi\right)R_{2}\left(0,\pi\right)R_{1}\left(\frac{\pi}{2},\frac{\pi}{2}\right)$ & $R_{3}\left(\frac{\pi}{2},\frac{\pi}{2}\right)R_{2}\left(0,\pi\right)R_{1}\left(0,\pi\right)$ \\
        $R_{3}\left(0,\pi\right)R_{2}\left(0,\pi\right)R_{1}\left(0,\pi\right)$ & $R_{3}\left(0,\pi\right)R_{2}\left(0,\pi\right)R_{1}\left(0,\pi\right)$ \\ \hline \hline
    \end{tabular}
    \caption{Sets of pre-rotation ($\mathcal{S}$) and post-rotation ($\mathcal{T}$) gates to implement quantum state tomography and quantum process tomography.}
    \label{tab:pre-post}
\end{table}
\endgroup

\begin{figure}[htbp]
    \centering
    \includegraphics[width=\linewidth]{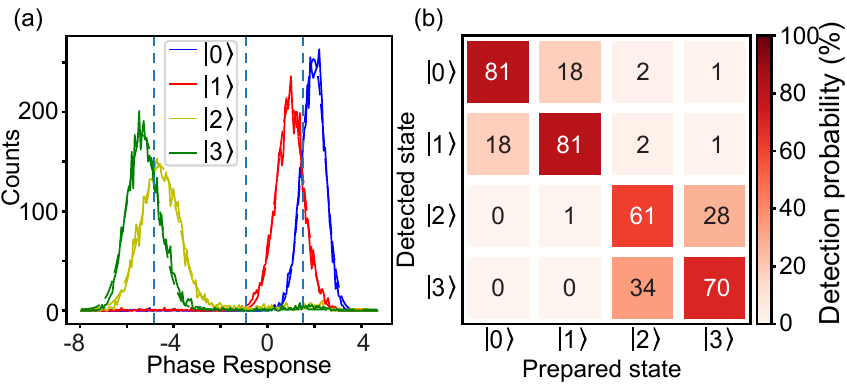}
    \caption{Readout characterization (a) Single-shot histogram after phase rotation and (b) heatmap. Single-ququart basis states are prepared and measured 5000 times.}
    \centering
    \label{fig:heatmap}
\end{figure}

The single-shot readouts are collected after applying the post-rotations. The readout histogram and heatmap of the assignment fidelity matrix are visualized in Figure \ref{fig:heatmap}. We apply the confusion matrix, which is the inverse matrix of the normalized readout heatmap, to the single-shot data to reduce the readout error. Maximum Likelihood Estimation (MLE) is then used to reconstruct the physical density matrix.

The single-ququart process tomography (QPT)~\cite{chuang_prescription_1997} is performed by sandwiching the unknown quantum channel $\mathcal{E}$ with different combinations of pre-rotation and post-rotation gates, where the pre-rotation set $\mathcal{S}$ is shown in Table \ref{tab:pre-post}.
% The pre-rotation set $T$ is:  $T=\{I, R_{01}\left(0,\frac{\pi}{2}\right), R_{01}\left(\frac{\pi}{2},\frac{\pi}{2}\right),R_{01}\left(0,\pi\right),R_{12}\left(0,\frac{\pi}{2}\right),$\\
% $R_{12}\left(\frac{\pi}{2},\frac{\pi}{2}\right),R_{12}\left(0,\pi\right)R_{01}\left(0,\frac{\pi}{2}\right),R_{12}\left(0,\pi\right)R_{01}\left(\frac{\pi}{2},\frac{\pi}{2}\right),$\\
% $R_{12}\left(0,\pi\right)R_{01}\left(0,\pi\right), R_{23}\left(0,\frac{\pi}{2}\right), R_{23}\left(\frac{\pi}{2},\frac{\pi}{2}\right),$\\ $R_{23}\left(0,\pi\right)R_{12}\left(0,\pi\right)R_{01}\left(0,\pi\right),$\\
% $R_{23}\left(0,\pi\right)R_{12}\left(0,\pi\right)R_{01}\left(0,\frac{\pi}{2}\right),$\\
% $R_{23}\left(0,\pi\right)R_{12}\left(0,\pi\right)R_{01}\left(\frac{\pi}{2},\frac{\pi}{2}\right),$\\
% $R_{23}\left(0,\pi\right)R_{12}\left(0,\frac{\pi}{2}\right), R_{23}\left(0,\pi\right)R_{12}\left(\frac{\pi}{2},\frac{\pi}{2}\right)\}$.

After applying each pre-rotation and the gate for characteristics, the density matrix is reconstructed through MLE QST. The full quantum channel $\mathcal{E}(\rho) = \sum_{i,j} \chi_{ij} B_i \rho B_j^\dagger$ is calculated based on Ref.~\cite{QPT2008} with either the basis $\qty{B_i} = \{Z_4^{m}X_4^{n}\}$ (ququart scenario) or $\qty{B_i} = \{\sigma_m \otimes \sigma_n\}$ (two-qubit scenario, ququart encoding), $m,n=\{0,1,2,3\}$. Here $Z_4$ and $X_4$ are the single-ququart Pauli operators and $\sigma_m$ denotes one of the qubit Pauli operators $\qty{\mathbb{1}, X, Y, Z}$. The operator basis elements are orthogonal under the Hilbert-Schmidt inner product: $\Tr[B_i^\dag B_j] = d \, \delta_{ij}$. The process fidelity of two channels with process matrices $\chi_1$ and $\chi_2$ is then given by
\begin{align}
    F = \abs{\Tr[\chi_1 \chi_2]},
\end{align}
which in the case of unitary channels recovers the expression~\ref{eq: infidelity}. Table~\ref{tab:qpt_fid} shows our experimental QPT results.

\addtolength{\tabcolsep}{2pt}
\begin{table}[htbp]
	\begin{tabular}{|cccc|}
	    \hline
        \hline
	     Gate & DEC & \multicolumn{2}{c|}{QOC}  \\
      & & no VZ & with VZ \\ \hline\hline
       $\mathbb{1}_4$ & \multicolumn{3}{c|}{$89.88\%$} \\ \hline
       $Z_4$ & \multicolumn{3}{c|}{$88.85\%$} \\ \hline
          $H_4$ & $84.30\%$ & $81.35\%$ & $86.18\%$ \\ \hline
          $H\otimes H$ & $83.53\%$ & $84.06\%$ & $-^*$  \\ \hline
          $X\otimes H$ & $83.47\%$ & $-$ & $-$ \\ \hline
          $H\otimes \mathbb{1}$ & $84.80\%$ & $-$ & $-$ \\ \hline
          $C\!X_{c=q_1}^{t=q_0}$ & $85.36\%$ & $-$ & $-$ \\ \hline\hline
	\end{tabular}
    \caption{Single-ququart gate fidelity characterized by QPT, for decomposed gates (DEC) and optimally controlled gates (QOC). In the QOC implementation of $H\otimes H$ we only considered the case without a final VZ, see Appendix~\ref{supp:qoc}.}
    \label{tab:qpt_fid}
\end{table}
\addtolength{\tabcolsep}{-2pt}

\section{Ququart RB}
\label{supp:rb}
In the RB protocol~\cite{magesan_scalable_2011} one constructs sequences of $m$ Clifford gates $C_m C_{m-1} \cdots C_1$ and further computes inversion gates $C_\mathrm{inv}$, which invert the effect of the sequences: $C_\mathrm{inv} C_m C_{m-1} \cdots C_1 = \mathbb{1}$. $C_\mathrm{inv}$ can be computed efficiently due to the Gottesman-Knill theorem~\cite{gottesman_the_1998}. In the experiment, one initializes the system in the ground state, applies the sequence, and measures the population in the ground state (survival probability) $P_m$.

Averaging over different sequences effectively corresponds to a depolarizing channel and the depolarization parameter $p$ can be extracted from a fit to the exponential decay $\langle P_m \rangle = Ap^m + B$. The error per Clifford gate (or for a specific Clifford gate in the IRB setting) is then determined as described in Section \ref{sec:rb}.

Since RB relies on sampling from the Clifford group, choosing the appropriate Clifford group for a certain scenario is important. The Clifford group $\mathcal{C}$ is defined as the group which normalizes the Pauli group $\mathcal{P}$. In the ququart case, the Pauli group $\mathcal{P}_4$ consists of products of $X_4$ and $Z_4$, which are normalized by $H_4$ and $S_4 = \mathrm{diag}\qty(1, \sqrt{i}, i, \sqrt{i})$~\cite{hostens_stabilizer_2005}. $Z_4$ trivially is an element of the ququart Clifford group $\mathcal{C}_4$ and we check numerically that $\qty{H_4, Z_4, S_4}$ can generate the predicted number of 768 elements in $\mathcal{C}_4$ \cite{tolar_on_2018}. In the case of two qubits, the Pauli group $\mathcal{P}^{\otimes 2}_2$ is given by the tensor product of the single-qubit operators $\mathbb{1}$, $X$, $Y$ and $Z$, and the gates $\qty{H, S, C\!X}$ generate the Clifford group $\mathcal{C}^{\otimes 2}_2$ with 11520 elements (up to scalars)~\cite{selinger_generators_2015}.

\addtolength{\tabcolsep}{2pt}
\begin{table}[t]
    \begin{tabular}{|ccc|}
	    \hline
        \hline
	     Gate & DEC & QOC \\  \hline\hline
          $H_4$ & $96.18(32)\%$ & $96.41(32)\%$ \\ \hline
          $H\otimes H$ & $95.17(19)\%$ & $95.98(19)\%$ \\ \hline
          $\mathbb{1}\otimes H$ & $98.51(13)\%$ & $-$ \\ \hline
          $H\otimes \mathbb{1}$ & $96.70(29)\%$ & $-$ \\ \hline
          $C\!X_{c=q_0}^{t=q_1}$ & $96.68(24)\%$ & $-$ \\ \hline
          $C\!X_{c=q_1}^{t=q_0}$ & $98.84(17)\%$ & $-$ \\ \hline\hline
	\end{tabular}
    \caption{Single-ququart gate fidelity characterized by different types of RB experiments, for decomposed gates (DEC) and optimally controlled gates (QOC). The $H_4$ gate fidelity is characterized using the Clifford group $\mathcal{C}_4$, other gates are characterized using the Clifford group $\mathcal{C}^{\otimes 2}_2$. Base fidelity for RB with $\mathcal{C}_4$ is $96.22(14)\%$, and with $\mathcal{C}^{\otimes 2}_2$ is $95.84(5)\%$. We only consider the VZ implementation for the QOC $H_4$ pulse.}
	\label{tab:rb_fid}
\end{table}
\addtolength{\tabcolsep}{-2pt}

For the ququart case, we sample from $\mathcal{C}_4$ which we construct explicitly, and for the two-qubit case we use Qiskit~\cite{qiskitcontributors_qiskit:_2023} to produce RB sequences. In either case, we decompose the matrices for each gate in a sequence according to decomposition~\eqref{eq:decomposition} in order to implement the RB experiment on our transmon. RB and IRB results are summarized in Table~\ref{tab:rb_fid}.

\section{Error budget}
\label{supp:sim}
Table~\ref{tab:error_budget} shows the infidelities of different ququart gates extracted from simulating the Lindblad master equation. We compare the noise-free case with the noisy case, where for the latter we include decay and dephasing collapse operators for each subspace using the experimentally measured lifetimes. The simulation time step is set to $\dd{t} = \SI{0.001}{\nano\second}$ for high numerical precision, which is over an order of magnitude lower than what was used for the pulse optimization (see Appendix~\ref{supp:qoc}). We perform all simulations with the lowest $5$ energy levels of the transmon to account for a leakage into higher levels. By taking the trace of the $4 \times 4$ submatrix which describes the ququart state, we verify for each simulation result that leakage into the fifth state is below $0.01\%$, justifying the truncation of the Hamiltonian at $\tilde{d}=5$ levels.  

We observe that the optimal control pulses can theoretically reduce the decoherence-free infidelities by more than $30\%$, and future work will be optimizing pulse duration to reduce the decoherence error further. The difference in infidelity between the ideal simulation here and the optimization target $1-F_{\vec{\phi}} \leq 10^{-4}$ arises from the increased time step precision.

\begin{table}[htbp]
	\begin{tabular}{|ccc|}
	    \hline
        \hline
	     Gate & Decoherence-free & Decoherence  \\ \hline\hline
      $H_4$ DEC & $0.42\%$ & $4.53\%$ \\ \hline
      $H \otimes H$ DEC & $0.41\%$ & $4.78\%$ \\ \hline
      $C\!X_{c=q_0}^{t=q_1}$ DEC & $0.38\%$ & $4.11\%$ \\ \hline
      $C\!X_{c=q_1}^{t=q_0}$ DEC & $0.03\%$ & $2.18\%$\\ \hline
      $H \otimes I$ DEC & $0.62\%$ & $3.89\%$  \\ \hline
      $I \otimes H$ DEC & $0.11\%$ & $1.82\%$ \\ \hline
      $H_4$ QOC & $0.08\%$ & $4.28\%$ \\ \hline
      $H \otimes H$ QOC & $0.26\%$ & $4.81\%$  \\ \hline\hline
          
	\end{tabular}
    \caption{Ququart gate infidelities extracted from  simulation. DEC: Decomposition, QOC: Quantum optimal control.}
    \label{tab:error_budget}
\end{table}

\normalem{}
\newpage
\bibliography{bibliography}
\end{document}